# Realizing high-temperature superconductivity in compressed molecular-hydrogen through Li doping

Ashok K. Verma[*,1,2] and P. Modak[1,2]

[1]*High Pressure & Synchrotron Radiation Physics Division
Bhabha Atomic Research Centre, Mumbai, India -400085,*
[2]*Homi Bhabha National Institute
Anushaktinagar Mumbai, India- 400094*
[*]*E-mail address: hpps@barc.gov.in*

*Abstract*

In this study, we explore lithium-doped stable molecular hydrogen structures by performing first-principles crystal structure searches across varying compositions in the Li-H system under high pressure. Our search reveals a cubic phase of $LiH_{12}$, which shows promise as a high-temperature superconductor. Our Bader charge analysis suggests that electron transfer from Li to H atoms tunes the intra- and inter-molecular H-H distances, which are critical for the metallization of molecular hydrogen. This modulation alters the interaction between bonding and anti-bonding $1s$ states of hydrogen molecules. Furthermore, Li ions serve as stabilizers for the distorted $H_2$ molecular network through ionic interactions. Numerical solutions to the fully anisotropic Migdal-Eliashberg equations reveals that this phase could exhibit superconductivity above 300 K at a pressure of 250 GPa, a pressure value that is typically achievable using a diamond anvil cell. Detailed analysis of species-specific phonons and the Eliashberg function shows that low- and intermediate-energy phonons are crucial in promoting strong electron-phonon coupling. Thus, our study establishes lithium doping as a promising approach to induce high-temperature superconductivity in compressed molecular hydrogen without causing molecular dissociation.

The quest of room-temperature superconductivity dates back to Ashcroft's 1968 proposal, which suggested that compressed atomic hydrogen metal, could exhibit Bardeen-Cooper-Schrieffer (BCS) type superconductivity at elevated temperatures. This was attributed to its high phonon frequencies, large density of states at the Fermi level, and strong electron-phonon coupling [1]. In the early years, since atomic hydrogen metal was experimentally inaccessible, researcher efforts shifted towards exploring superconductivity in metal-hydrides that remained stable under ambient conditions. Many hydrides were scanned for superconductivity with limited success. $Th_4H_{15}$ was the first notable hydride which was discovered with a critical temperature ($T_C$) of ~8 K [2, 3]. However, as this program was not making expected progress, attention eventually shifted to cuprates, a new class of materials that demonstrated ambient pressure superconductivity at much higher temperatures, around 133 K [4, 5]. While some cuprates showed a rise in $T_C$ under high pressure (~ 164 K at 30 GPa) [6], none have exhibited a significant enhancement. Notably, despite extensive experimental efforts, the formation of metallic hydrogen has yet to be conclusively demonstrated, although potential signatures of its existence have been reported, even at pressures up to 495 GPa [7-10].

In 2004, attention of researcher once again turned to hydride superconductivity following a new proposal by Ashcroft, who suggested that high pressure could metallize hydrogen-rich alloys or hydrides that would otherwise be insulating under ambient conditions [11, 12]. In these materials, hydrogen molecules would exist in a "chemically precompressed" state, which was expected to facilitate the dissociation of $H_2$ molecules into atomic H-metal at much lower pressures than those required for pure hydrogen. Similar to metallic hydrogen, conventional electron-phonon coupling was anticipated to drive high-TC superconductivity in these hydrides. Interestingly, the concept of "chemically precompressed" hydrogen was first introduced by Gilman in 1970 [13], but it gained broader recognition and attention following Ashcroft's 2004 proposal.

Since Ashcroft's 2004 proposal, a variety of binary hydrides, including $H_3S$ [14], $LaH_{10}$ [15], $YH_6$ [15], $YH_{10}$ [15] $CaH_6$ [16], $ThH_{10}$ [17], $CeH_9$ [18], $AcH_{10}$ [19], and many others [20-22], have been predicted to be high-$T_C$ superconductors at pressures lower than those required for the formation of atomic hydrogen-metal. In line with the original proposal, all these high-$T_C$ materials are characterized by the presence of atomic hydrogen. Surprisingly only a limited number of predicted hydrides such as $H_3S$ [23], $LaH_{10}$ [24, 25], $YH_6$ [26], $YH_9$ [26] and $CaH_6$ [27] have been successfully confirmed in the diamond-anvil cell (DAC) experiments, to date. This highlights both the experimental challenges involved and the considerable efforts still required to achieve the long-sought goal of room-temperature superconductivity. Despite the challenges, these discoveries represent significant progress in the quest of room-temperature superconductivity, positioning hydrides at the forefront of high-$T_C$ race.

Despite extensive searches across a broad range of metal hydrides, only a limited number have been identified to exhibit superconductivity above room-temperature. These includes $Li_2MgH_{16}$ (473K at 250 GPa) [28], $Li_2NaH_{17}$ (340 K at 300GPa) [29], $Li_2NaH_{23}$ (310 K at 350 GPa) [29] and $LaSc_2H_{24}$ (331K at 250 GPa) [30], $Li_2CaH_{16}$ (330 K at 350 GPa) [31], $Li_2CaH_{16}$ (370 K at 300 GPa) [31] etc. Notably, all of the room-temperature superconductors are ternary hydrides, and to date, no binary hydride has been found to exhibit room-temperature superconductivity at any pressure. A distinguishing feature of these high-$T_C$ metal hydrides, as identified to date, is the formation of cage-like structures around the metal ions, with H-H bond lengths ranging from 1.0 to 1.5 Å, due to the partial/full dissociation of $H_2$ molecules. This structural feature is considered crucial for achieving high-$T_C$ superconductivity in hydrides [20-22]. While the pressures required for these materials are achievable with current diamond-anvil cell (DAC) devices, these predictions have yet to be experimentally validated. This highlights the pressing need for simulation groups to explore new room-temperature superconductors, thereby providing additional synthesis pathways for experimental research.

In this study, we use controlled electron doping to achieve high-$T_C$ superconductivity in a molecular hydrogen system under pressure. The introduction of additional electrons leads to a modest weakening of the H-H bond without causing molecular dissociation, thereby maintaining an H-H bond length significantly shorter than that found in atomic-H metal [32-34] and in known hydride superconductors [20-22]. This doping effect stabilizes the molecular hydrogen metal at lower pressures than those required for the metallization of pure molecular hydrogen [35]. Towards this, we focus on Li-doped molecular hydrogen as a test case, distinguishing our approach from previous studies that have examined electron doping in metallic superhydrides [15, 20-22]. Lithium was chosen as an electron donor due to its low atomic mass and low electronegativity, which facilitates the easy transfer of electrons from Li to H, an element with high electronegativity. Furthermore, synthesis of binary hydrides is expected to be easier than ternary or higher hydrides. Indeed, recent DAC experiments have successfully synthesized novel Li hydrides such as $LiH_2$ and $LiH_6$ [36], confirming the earlier predictions of H-rich hydride formation at pressures above 100 GPa [37].

Over the years, density functional theory (DFT)-based crystal structure searches have become the backbone of the hydride superconductivity research, uncovering numerous novel hydride superconductors [20-22]. Most of the currently known superconducting hydrides were first identified through such searches. In this work, we explore high pressure phase diagram of Li-H by conducting extensive variable composition ($N$ = 8-16, 10-20, 12-24 & 16-32) searches at 150, 250 and 350 GPa, using evolutionary algorithm as implemented in the USPEX code [38-40]. This method has been successfully applied to various materials under different thermodynamic conditions [41-44]. We first constructed the full Li-H convex hull to identify thermodynamically stable stoichiometries, and subsequently performed detailed electronic structure calculations to screen for potential high-$T_C$ superconductors. For the most promising phase, we solved the fully anisotropic Migdal-Eliashberg equations to obtain quantitative estimates of their superconducting transition temperatures. By employing this strategy, we uncovered a unique H-rich hydride, namely $LiH_{12}$, which, in its simple cubic phase, exhibits superconductivity with a $T_C$ above 300 K at 250 GPa.

Our ab-initio calculations employ the projector augmented wave (PAW) method [45] and the Perdew-Burke-Ernzerhof (PBE) exchange-correlation functional [46] as implemented in the Vienna Ab-initio Simulation Package (VASP) [47-49]. For structural optimization in the searches, a plane-wave basis set was utilized with a 500 eV energy cutoff and the Brillouin zone was sampled using the Monkhorst-Pack method [50] with a grid spacing of $2\pi \times 0.06$ Å$^{-1}$. For re-optimization of the most favorable structures, the plane-wave basis set energy cutoff was increased to 600 eV, while reciprocal space integrations were performed with a finer grid spacing of $2\pi \times 0.03$ Å$^{-1}$. Ionic positions were relaxed until the forces were reduced to less than 2.0 meV/Å. These computational parameters ensured the convergence of total energies to within 1.0 meV per atom. All electron frozen-core projector-augmented wave (PAW) potentials were employed with Li ($1s^2, 2s^1$), and H ($1s^1$) as the valence configurations. All crystal structures were visualized using VESTA software [51]. The zero-point phonon energy was calculated using quantum-espresso software with ultra-soft pseudopotentials [52]. To assess the thermal stability of cubic $LiH_{12}$, we performed finite-temperature ab initio molecular dynamics simulations in the canonical (NVT) ensemble, using a Nosé thermostat [53]. The simulations were performed using a 351-atom cubic supercell (27 Li and 324 H) with a time step of 0.5 fs. The volume was kept fixed to that of the 250 GPa value at 0K.

Our search revealed total 14 Li-H compositions on the convex hull which are stable against their dissociation into the mixtures of pure Li and H elements, as shown in the Fig.1a. Barring LiH, all other hydrides can be classified into two groups: Li-rich and H-rich. Since LiH is the sole stable hydride of Li at ambient conditions, we examined the stability of the Li-rich and H-rich compounds against dissociation into LiH, Li, and LiH, $H_2$ mixtures, respectively. While $Li_{10}H$, $Li_{10}H_3$, $Li_2H$ and $Li_5H_3$ Li-rich hydrides become metastable (as shown in the Fig. S1), all H-rich hydrides remained stable, as shown in the Fig. 1b.

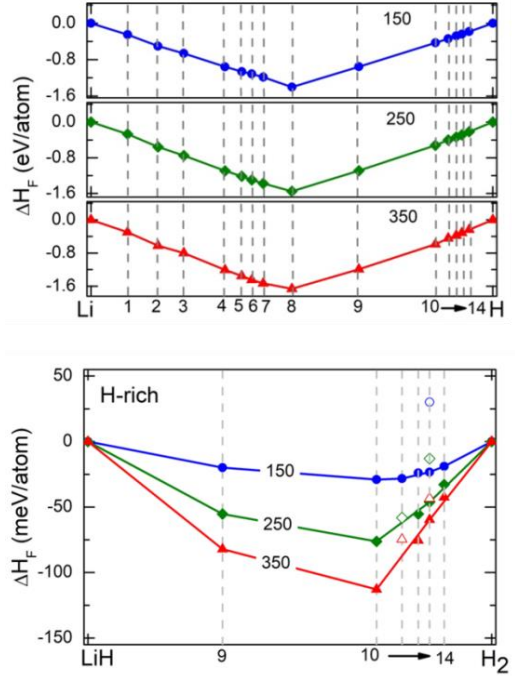

**FIG. 1:** DFT convex-hull for Li-H at high pressures: (*a*) Formation enthalpies of hydrides relative to the dissociation of products of mixtures of Li and H. (*b*) Formation enthalpies of H-rich hydrides relative to the dissociation of products of mixtures of LiH and $H_2$. Here, 1→ $Li_{10}H$, 2→ $Li_5H$, 3→$Li_{10}H_3$, 4→$Li_2H$, 5→$Li_5H_3$, 6→$Li_3H_2$, 7→$Li_4H_3$, 8→ LiH, 9→$LiH_2$, 10→ $LiH_6$, 11→$LiH_8$, 12→$LiH_{10}$, 13→ $LiH_{12}$ and 14→ $LiH_{16}$. Open symbols represent the metastable structures.

Notably, our searches revealed three novel hydrides —$LiH_{10}$, $LiH_{12}$ and $LiH_{16}$— along with $LiH_2$, $LiH_6$ and $LiH_8$ which were also identified in earlier study [37]. Analysis of internal-energy (ΔU) and pressure-volume (PΔV) components of the enthalpy indicates that pressure favors hydride formation, as shown in the Fig. S1. Structural information for hydrides is provided in Tables S1 & S2. Li-rich compounds contain only atomic H, while H-rich compounds (except for $LiH_2$, which has both H and $H_2$ units) contain $H_2$ molecules, as seen in Fig. S2. The phase diagram for all compounds was computed in the 150-350

GPa pressure range, as shown in Fig. S3. It is important to note that zero-point phonon energies were not included in the enthalpies of the convex hull diagrams, as their contributions are small (1-7 meV/atom) according to previous studies [37]. We are therefore confident that zero-point phonon energy contributions will have a minimal impact on our conclusions. This is further supported by a recent study, which through detailed analysis of finite temperature effects and DFT inaccuracies, suggested a significantly larger enthalpy window of 50 meV/atom above the DFT-calculated convex hull for the thermodynamic stability of the compounds [54]. Notably, nearly 80% of the inorganic compounds in the ICSD database lie above the DFT convex hull [54, 55]. Therefore we expect that all 14 Li-H compounds could be synthesized in DAC experiments under favorable conditions.

We examined the electronic properties of Li-hydrides by calculating the electronic density of states, Bader charges and electron-localization function (ELF) at 250 GPa. Apart from $Li_5H$, all hydrides are metallic, as indicated by the presence of finite DOS at the Fermi level ($E_F$), as shown in the Figs. S4 & S5. A comparison of $E_F$ DOS values for H-rich hydrides, as displayed in the Fig. 2, reveals that a cubic phase of $LiH_{12}$ has an $E_F$ DOS comparable to that of the atomic H metal whereas the $E_F$ DOS values of other compounds are considerably lower than that of the cubic $LiH_{12}$. Furthermore, we analyzed the chemical bonding in H-rich hydrides by computing the Bader charges, which reveals a charge transfer from Li to the antibonding state of the $H_2$ molecules, as given in the Table 1. Thus chemical bonding between Li and $H_2$ has an ionic component also. The ELF further confirms the presence of ionic interactions between Li and $H/H_2$ units, as closed ELF isosurfaces were seen around atomic sites (as shown in the Fig. S6). The bolster-like ELF isosurface around $H_2$ molecules indicates presence of covalent interaction between H-H atoms.

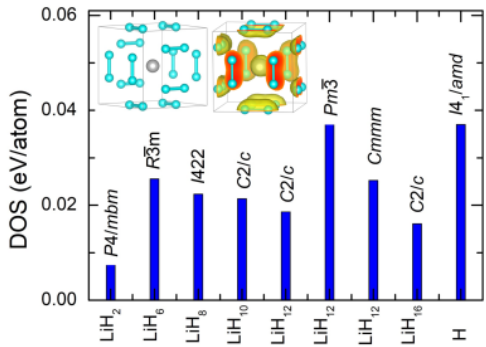

**FIG. 2:** Comparison of the Fermi-level DOS values for H-rich compounds along with atomic H metal at 250 GPa.

The Fermi level DOS of cubic $LiH_{12}$ (space group: $Pm\bar{3}$) is marginally higher than that of the atomic H-metal. Inset depicts the cubic $LiH_{12}$ lattice and ELF isosurfaces for isovalue = 0.85. Here, grey ball displays Li atom and small cyan balls display H atoms.

In view of the strong correlation between the electron-phonon coupling constant and $E_F$ DOS values [56], cubic $LiH_{12}$ emerges as a particularly promising candidate for high-temperature superconductivity among H-rich hydrides. This can be attributed to its high $E_F$ DOS value and high-symmetry crystal structure. Indeed, several high-temperature superconducting hydrides, including $H_3S$, $LaH_{10}$, and $CaH_6$, are known to possess these two key characteristics [20-22]. Additionally, Mathias' rules further support the idea that a high DOS at the Fermi level and a high-symmetry cubic structure are favorable for superconductivity [57]. It is also worth noting that a previous study has predicted superconductivity in $LiH_6$ ($T_C$ > 50 K at 250 GPa) and $LiH_8$ ($T_C$ > 35 K at 200 GPa) under high pressure conditions [58]. More recently, superconductivity was explored in other Li-H compounds, such as $LiH_4$ and $Li_8H_n$ ($n$ = 4–7) as well [59–61].

**TABLE 1:** Calculated H-H distances and Bader charge transfer from Li to H atoms at 250 GPa.

| Systems (Space-group) | H-H Bond-length (in Å) | Bader Charges (in $e$) |
|---|---|---|
| Molecular H ($C2/c$) | 0.75 | --- |
| LiH ($Fm\bar{3}m$) | --- | Li: +0.72<br>H: -0.72 |
| $LiH_2$ ($P4/mbm$) | 1.61<br>0.75 | Li: +0.76<br>H1: -0.65<br>H2: -0.11 |
| $LiH_6$ ($R\bar{3}m$) | 0.82 | Li: +0.79<br>H: -0.13 |
| $LiH_8$ ($I422$) | 0.80 | Li: +0.79<br>H: -0.10 |
| $LiH_{10}$ ($C2/c$) | 0.78<br>0.80<br>0.83 | Li: +0.79<br>H: -0.04 to -0.11 |
| $LiH_{12}$ ($C2/c$) | 0.78<br>0.79<br>0.80 | Li: +0.80<br>H: -0.03 to -0.08 |
| $LiH_{12}$ ($Pm\bar{3}$) | 0.83 | Li: +0.82<br>H: -0.07 |
| $LiH_{16}$ ($C2/c$) | 0.76<br>0.77<br>0.78 | Li: +0.80<br>H: -0.03 to -0.05 |

| atomic H-metal ($I4_1/amd$) | 1.05 | --- |

A detailed analysis of electronic band structure and partial DOSs, as shown in Fig. 3, reveals that the $E_F$ DOS value of cubic LiH$_{12}$ is primarily dominated by the H-$1s$ orbitals, a characteristics shared by several high-$T_C$ hydride superconductors [20-22]. This observation is further supported by the band decomposed charge densities associated with the filled bands 1, 2, and 3, which surround the hydrogen atoms, as illustrated in the Fig. S7. The striking similarity in the electronic properties of the cubic LiH$_{12}$ and its H sub-lattice, as depicted in the Fig. 3, indicates that cubic LiH$_{12}$ can be considered as electron-doped molecular hydrogen. A similar trend is also observed in the electronic properties of other hydrogen-rich compounds, as illustrated in the Fig. S8. Since pure molecular hydrogen exhibits insulating behavior under this pressure [32, 35], the metallic nature of H sub-lattices appears to be closely associated with distortions within the molecular hydrogen network [62, 63]. In contrast to molecular hydrogen, which forms a layered structure resembling distorted graphene (with distortions in the 6H rings) [32], the hydrogen sublattice in cubic LiH$_{12}$ exhibits significantly greater distortion, as shown in the Fig. S3. These distortions are also likely responsible for the emergence of two prominent features in the DOS function of the H sublattices: a pseudogap and a van Hove singularity near the Fermi level. An enthalpy comparison of hydrogen sub-lattices with that of pure molecular hydrogen further strengthens the distortion picture. Our results show that upon full relaxation cubic H sub-lattice is ~228 meV/H less stable than the $C2/c$-H [32] at 250 GPa. It is also noteworthy that the fully relaxed H sub-lattice of $C2/c$ LiH$_{12}$ is simply ~11 meV/H less stable than $C2/c$-H. In contrast to this, H sub-lattice of LaH$_{10}$ is ~550 meV/H less stable than $C2/c$-H.

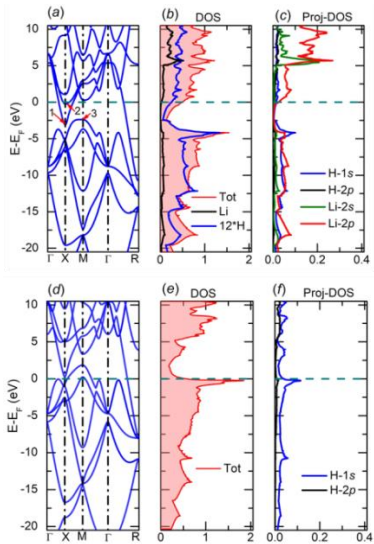

**FIG. 3:** Electronic properties of the cubic LiH$_{12}$ (*upper panel*) and H sub-lattice (*lower panel*) at 250 GPa. (*a*) Electronic band structure of cubic LiH$_{12}$ along high symmetry directions of Brillouin zone. (*b*) The total electronic density of states (DOS) projected onto Li and H atoms. (*c*) The projected DOS onto H-$1s$, H-$2p$, Li-$2s$ and Li-$2p$ orbitals. Dark cyan dashed horizontal line indicates the Fermi level. The same nomenclature is also applied in the lower panel.

To understand the covalent bonding of the H-H atoms, we calculated the COHP, ICOHP and ICOBI, as illustrated in Table 2 [64, 65]. This analysis reveals that the H-H bonds in cubic LiH$_{12}$ are very strong having covalent character around 47%, while the Li-H interactions are characterized by significantly weaker bonds (ICOHP = -0.69 eV/pair & ICOBI = 4.7%). The H-H pair in LiH$_6$ and LiH$_8$ also displays similar characteristics. The Li-H pair in other hydrides exhibits similar bond characteristics (see Table S5). In high-$T_C$ superconducting materials such as Li$_2$MgH$_{16}$ and LaH$_{10}$, the strength of H-H covalent bond is significantly weaker, leading to longer bonds [15, 28]. Therefore, this analysis suggests that Li hydrides are more similar to molecular hydrogen than to atomic hydrogen metal.

**TABLE 2:** Calculated covalent bond parameters of H-H pair at 250 GPa. Integrated crystal orbital Hamiltonian population (ICOHP) and integrated crystal orbital bond indices (ICOBI) quantify covalent bond strength and character, respectively. For this analysis, only the nearest-neighbor pairs are considered.

| Systems (Space-group) | ICOHP (in eV/pair) | ICOBI (in %) |
|---|---|---|
| H molecular ($C2/c$) | -6.20 | 60 |
| LiH$_{12}$ ($Pm\bar{3}$) | -5.27 | 47 |
| LiH$_6$ ($R\bar{3}m$) | -4.97 | 48 |
| LiH$_8$ ($I422$) | -5.43 | 47 |
| Li$_2$MgH$_{16}$ ($Fm\bar{3}m$) | -2.39 | 21 |
| LaH$_{10}$ ($Fm\bar{3}m$) | -1.36 | 12 |
| atomic H-metal ($I4_1/amd$) | -1.75 | 16 |

Interestingly the cubic phase of LiH$_{12}$ does not lie on the static DFT convex hull, as shown in the Fig1b. At

150 GPa, the enthalpy of the cubic phase is 54 meV/atom above the convex hull, but at 250 GPa, it decreases to 40 meV/atom, entering the proposed 50 meV/atom enthalpy window for thermodynamic stability [54]. Figure 4 displays the static enthalpies of four distinct phases of LiH$_{12}$. This phase becomes the lowest-energy structure at approximately 650 GPa, suggesting its remnant metastability at lower pressures [55]. Consequently, this phase may be synthesized at lower pressures [55]. Incorporating harmonic zero-point phonon energy reduces the transition pressure to 325 GPa. We anticipate that the transition pressure will further decrease when anharmonic phonon contributions are considered, due to the low atomic masses of Li and H [66]. Furthermore, studies on LaH$_{10}$, including full nuclear quantum effects, suggest that low-symmetry structures of LiH$_{12}$ may not remain stable when fully accounting for nuclear quantum fluctuations [67]. Our phonon calculations reveal that cubic phase has stable phonon modes at pressures as low as 200 GPa. Thus, the synthesis of this phase could be feasible at 200 GPa in DAC experiments, owing to its remnant metastability [55].

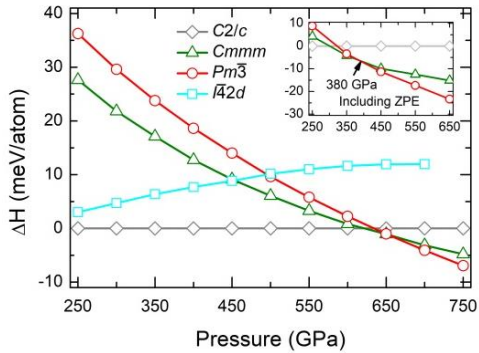

**FIG. 4:** Calculated enthalpy as a function of pressure for the LiH$_{12}$ phases relative to the *C*2/*c* phase. Inset: Calculated enthalpy, including harmonic phonon zero-point energy, as a function of pressure.

We investigated the thermal stability of the cubic LiH$_{12}$ phase by performing molecular dynamics simulations at various temperatures, with the results presented in Figure 5. These simulations indicate that the Li sublattice remains stable up to temperatures as high as 600 K. However, the H sublattice begins to exhibit liquid-like diffusion behavior at temperatures above 300 K.

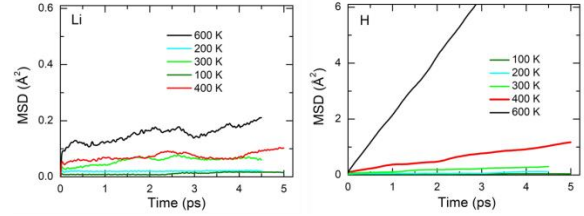

**FIG. 5:** Mean square displacements (MSDs) of Li and H atoms in the cubic phase of LiH$_{12}$, calculated at a fixed volume corresponding to the 250 GPa at 0K, at temperatures of 100, 200, 300, 400, and 600 K.

Based on the preceding discussion we identified cubic LiH$_{12}$ as a potential high-$T_C$ superconductor. Therefore we calculated phonon properties and electron-phonon interactions of cubic LiH$_{12}$ using density functional perturbation theory [52, 68]. By calculating the electronic spectrum, phonon spectrum, and their interactions, we solved fully anisotropic Migdal-Eliashberg equations using the Wannier interpolation technique, as implemented in the Quantum-Espresso [69-71] and EPW codes [72–74]. The details of this set of calculations are given in Ref. [75]. Figure 6 shows the superconducting gap distribution as a function of temperature for a typical coulomb pseudopotential value $\mu^* = 0.10$. Our calculated $T_C$ is around 400 K at 250 GPa which reduces around 340 K for $\mu^* = 0.20$. It is obvious that cubic LiH$_{12}$ is a single gap superconductor having a gap value of ~ 88 meV at 25 K.

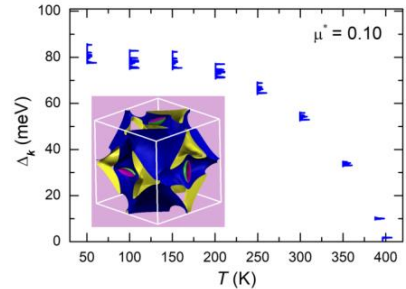

**FIG 6:** Calculated anisotropic superconducting gap $\Delta_\mathbf{k}$ of cubic LiH$_{12}$ on the Fermi surface as a function of temperature. The inset shows the 3*D* Fermi surface.

Figure.7 depicts the phonon dispersion, phonon density of states, isotropic Eliashberg spectral function $\alpha^2F(\omega)$ and the increasing electron-phonon coupling constant $\lambda(\omega)$ which was obtained by integrating $\alpha^2F(\omega)/\omega$ over $\omega$. EPW code gives a converged value of 2.83 for electron-phonon coupling constant, while a value of 2.71 was obtained from the QE with a Gaussian smearing width of 0.005 Ry. It is also noteworthy that the Allen-Dynes-modified McMillan equation [56] results a significantly lower $T_C$ of ~202 K, for a coulomb psuedopotential $\mu^* = 0.10$ & a logarithmic average phonon frequency $\omega_{log}$ ~1175K, highlighting the importance of fully anisotropic calculations for such materials. Furthermore, we observed that low frequency phonon modes, lying below 100 meV, contributes approximately 40% to overall $\lambda$ value and surprisingly the highest frequency phonon modes, lying above 330 meV, contributes less than 10% to overall $\lambda$ value. This observation aligns well with the fact that $\lambda \propto \omega^{-2}$. A similar trend has also been reported in several other hydride superconductors [76]. It is important to note that our study does not incorporate the influence of quantum nuclear effects, which may significantly affect the phonon and superconducting properties, as indicated by recent studies on many H-rich hydride superconductors [20, 66, 67, 77 and 78]. However, a recent detailed analysis of quantum nuclear effects in various hydride superconductors reveals that structures having atoms in symmetric chemical environments are not only less impacted by these perturbations but also exhibit increased superconducting transition temperatures when these effects are taken into account [76]. Given the highly symmetric crystal structure of cubic $LiH_{12}$, we therefore believe that, although these effects present significant computational challenges, they are unlikely to fundamentally change the key conclusions regarding its superconducting properties.

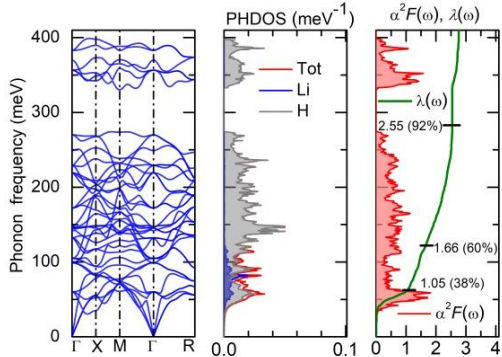

**FIG. 7:** Phonon dispersion along the high symmetry directions of Brillouin zone. (b) Total and atom projected phonon density of states (PHDOS) onto Li and H atoms. (c) Isotropic Eliashberg spectral function, $\alpha^2F(\omega)$, and running frequency dependent electron-phonon coupling strength $\lambda(\omega)$. Horizontal bar indicates the running total value of $\lambda(\omega)$.

In summary, we have achieved high-temperature superconductivity in molecular hydrogen at significantly lower pressures than those required for pure hydrogen by applying controlled Li doping. Through comprehensive first-principles crystal structure searches across various compositions in the Li-H system, we have uncovered a cubic phase of $LiH_{12}$ that exhibits superconductivity above 300 K at a pressure of 250 GPa, a pressure value attainable in diamond anvil cell experiments. This pressure could be further reduced by carefully selecting an appropriate metal dopant. Our detailed analysis of the electronic properties has revealed that hydrogen atoms make a dominant contribution to the total density of states at the Fermi level, playing a key role in the high-temperature superconductivity. The Li atoms, through electron transfer, stabilize the distorted $H_2$ molecular network. Further, investigation of species-specific phonons and the Eliashberg function has shown that low- and intermediate-energy phonons drive strong electron-phonon coupling in cubic $LiH_{12}$, while the contribution of the highest energy-phonons to the total $\lambda$ value is approximately 10%. Therefore, cubic $LiH_{12}$ emerges as the only binary hydride exhibiting superconductivity at temperatures exceeding room temperature.

**Acknowledgement:** The authors express their sincere gratitude to Dr. S. M. Yusuf and Dr. T. Sakuntala for their invaluable support throughout the execution of this project. The use of the ANUPAM Supercomputing Facility at BARC is also gratefully acknowledged.